# Common Occurrence of Explosive Hydrogen Burning in Type II Supernovae

Nan Liu[1,8 *], Thomas Stephan[2,3], Patrick Boehnke[2,3], Larry R. Nittler[1]
Bradley S. Meyer[4], Conel M. O'D. Alexander[1], Andrew M. Davis[2,3,5],
Reto Trappitsch[2,3,6], Michael J. Pellin[2,3,5,7]

[1]Department of Terrestrial Magnetism, Carnegie Institution for Science, Washington, DC 20015, USA

[2]Department of the Geophysical Sciences, The University of Chicago, Chicago, IL 60637, USA

[3]Chicago Center for Cosmochemistry, Chicago, IL, USA

[4]Department of Physics and Astronomy, Clemson University, Clemson, SC 29634, USA

[5]The Enrico Fermi Institute, The University of Chicago, Chicago, IL 60637, USA

[6]Nuclear and Chemical Sciences Division, Lawrence Livermore National Laboratory, Livermore, CA 94550, USA

[7]Materials Science Division, Argonne National Laboratory, Argonne, IL 60439, USA

## ABSTRACT

We report Mo isotopic data for 16 $^{15}$N-rich presolar SiC grains of type AB ($^{14}$N/$^{15}$N < solar, AB1) and their correlated Sr and Ba isotope ratios when available. Eight of the 16 AB1 grains show *s*-process Mo isotopic compositions, together with *s*-process Ba and/or Sr isotopic compositions. We found that a higher percentage of AB1 grains show anomalous isotopic compositions than that of AB2 grains ($^{14}$N/$^{15}$N > solar), thus providing further support to the division of the two AB subgroups recently proposed by Liu et al. (2017a), who showed that AB1 grains most likely originated from Type II supernovae (SNe) with explosive H burning. Comparison of the Sr, Mo, and Ba isotopic compositions of the AB1 grains with SN model predictions indicates that the *s*-process isotopic compositions of AB1 grains resulted from neutron-capture processes occurring during the progenitor massive stars' pre-SN evolution rather than from an explosive neutron-capture process. In addition, the observations of (1) explosive H burning occurring in the C-rich regions of the progenitor SNe of SN X grains as suggested by the isotopic compositions of X grains, and (2) explosive H burning occurring both at the bottom of the He/C zone and at the top of the He/N zone as suggested by model simulations, imply that explosive H burning is a common phenomenon in outer SN zones.

## 1. INTRODUCTION

Presolar grains are condensates formed in dying stars prior to solar system formation. They were ejected into the interstellar medium by stellar winds or explosions, traveled to the nascent solar system, survived destruction in the protosolar nebula, and were eventually incorporated into the parent bodies of primitive meteorites that formed within the first few million years of solar system history. As ancient stellar relicts, presolar grains record the isotopic compositions of their progenitor stars that resulted from Galactic chemical evolution and stellar nucleosynthesis (e.g., Nittler & Ciesla 2016). Diverse presolar mineral phases have been identified, and, of these, micrometer-sized presolar SiC grains have been most extensively studied for their chemical and isotopic compositions and crystal structures (e.g., Zinner 2014; Daulton et al. 2003). Presolar SiC grains are divided into five main groups (each group comprises ≥1% of presolar SiC), MS (mainstream), Y, Z, X, and AB, based on their C, N, and Si isotopic systematics. Extensive isotopic studies have shown that MS, Y, and Z grains came from low-mass AGB stars with near-solar, ~1/2 solar, and ~1/3 solar metallicities, respectively (e.g., Zinner 2014). On the other hand, X grains, which make up 1−2% of presolar SiC, have long been recognized to originate from Type II supernovae (SNe), based largely on their $^{28}$Si and $^{44}$Ti excesses (Nittler et al. 1996). The Si isotopic ratios of X grains span a large range, with the majority of the grains, the so-called X1 grains, lying along a single line in a plot of $\delta^{29}$Si[9] versus $\delta^{30}$Si with a slope of 0.67 (Nittler & Alexander 2003; Lin et al. 2010). In contrast, X2 grains show deviations from the 0.67 line to more negative $\delta^{29}$Si values. The large range of Si isotope ratios observed in X grains most likely reflects mixing between an inner zone enriched in $^{28}$Si from O burning and outer zones enriched in C, $^{29}$Si, and $^{30}$Si from He burning. An additional minor (<0.1%) group of presolar SiC, known as C grains, are also likely from supernovae. They are characterized by higher-than-solar $^{15}$N/$^{14}$N ratios, $^{28}$Si depletions, and a wide range of $^{12}$C/$^{13}$C ratios. Type C grains have been further divided into C1 and C2 grains; the latter have $^{12}$C/$^{13}$C ratios smaller than 10 (Liu et al. 2016).

[8]Present Address: Laboratory for Space Sciences and Physics Department, Washington University in St. Louis, St. Louis, MO, 63130, USA.

AB grains are a group of presolar SiC grains with particularly large $^{13}$C enrichments ($^{12}$C/$^{13}$C $\lesssim$ 10). They are relatively common (5−6%) among presolar SiC grains, but have the most ambiguous stellar origins amongst the five main SiC groups. A recent study of the light element isotopic compositions of AB grains revealed systematic differences in their C, N, and Si isotope ratios and initial $^{26}$Al/$^{27}$Al ratios between $^{15}$N-rich ($^{14}$N/$^{15}$N < solar, AB1)[10] and $^{14}$N-rich ($^{14}$N/$^{15}$N > solar, AB2) AB grains, strongly pointing to different stellar origins of these two subgroups (Liu et al. 2017a). The coexistence of explosive H burning and neutron-capture (*n*-capture) isotopic signatures in AB1 grains likely indicates an origin in SNe with explosive H burning (Liu et al. 2017a), and the contribution of SNe to the presolar SiC dust reservoir in the early solar nebula was therefore significantly higher than previously recognized, because AB1 grains (2−3%) are equally or more abundant than X grains (1−2%). On the other hand, AB2 grains do not show the negative correlation between $^{14}$N/$^{15}$N and initial $^{26}$Al/$^{27}$Al ratios (inferred from $^{26}$Mg excesses) observed in AB1 grains, and they generally show smaller $^{30}$Si excesses compared to AB1 grains (Liu et al. 2017a). A subsequent study found that AB2 grains have near-solar isotopic compositions of Sr, Mo, and Ba (Liu et al. 2017b), indicating that a majority of AB2 grains came from common but poorly understood stellar objects, J-type carbon stars, based on their shared characteristics: (1) similar $^{13}$C and $^{14}$N enrichments, (2) near-solar metallicities, and (3) lack of *s*-process isotopic signatures.

Literature data on the heavy-element isotopic signatures of AB1 grains are extremely sparse. Savina et al. (2003) studied Mo isotopic compositions of seven AB grains, with five of the seven being of type AB1, and found normal isotope ratios in four of them except for one grain with an unusual isotopic pattern exhibiting large excesses in $^{97,98}$Mo. On the other hand, Amari et al. (1995, 2001) observed trace-element abundance patterns with *s*-process enhancements in 13 of 20 AB1 grains measured in their study, in contradiction with the results of Savina et al. (2003). As a result, it is unclear whether or not AB1 grains inherited *s*-process isotopic signatures from their parent stars. Here, we report new Mo isotope data for 16 AB1 grains along with their Sr and Ba isotope data when available. The isotopic signatures of these AB1 grains are compared to a new set of 25 $M_\odot$ SN models, in which we investigate the effect of a large range of explosion energies on SN nucleosynthesis.

[9] $\delta^{i}$Si is defined as $\delta^{i}\mathrm{Si}=[(^{i}\mathrm{Si}/^{28}\mathrm{Si})_{grain}/(^{i}\mathrm{Si}/^{28}\mathrm{Si})_{std}-1]\times1000$, where $(^{i}\mathrm{Si}/^{28}\mathrm{Si})_{grain}$ and $(^{i}\mathrm{Si}/^{28}\mathrm{Si})_{std}$ represent the corresponding isotope ratios measured in a sample and the standard, respectively.

[10] The N isotope ratio of the protosun was determined to be 441±5 by Marty et al. (2011).

## 2. METHODS

The Sr, Mo, and Ba isotope data reported here (Table 1) were obtained in the same analytical sessions as those of 12 AB2 and 15 mainstream (MS) grains reported by Liu et al. (2017b) using the Chicago Instrument for Laser Ionization (CHILI; Stephan et al. 2016). The analytical details are given by Stephan et al. (2017). The C, N, Si, Al, S, and Ti isotope data of the AB1 grains, measured with the NanoSIMS 50L ion microprobe at the Carnegie Institution, were previously reported in Liu et al. (2017a), and their C, N, and Si isotope data are included in Table 1 for reference. In addition, the Raman spectral and electron dispersive X-ray features of some of the grains can be found in Liu et al. (2017c).

## 3. RESULTS AND SUPERNOVA MODELS

Figures 1a,b compare Mo isotopic data of the 16 AB1 grains with those of MS grains from Liu et al. (2017b). Eight of the 16 AB1 grains show *s*-process Mo isotopic compositions, which cover a similar range but shift more towards the solar composition relative to MS grains (Figs. 1a,b). Thus, in contrast to Savina et al. (2003), our study shows that a significant fraction of AB1 grains contain Mo with *s*-process, rather than solar, isotopic composition. Therefore, the predominance of isotopically normal Mo in the AB1 grains studied by Savina et al. (2003) probably indicates severe meteorite parent body and/or terrestrial Mo contamination sampled during their analysis. This is supported by the fact that 14 of 16 MS grains measured on the same mount by Savina et al. had $\delta^{92}$Mo values above −200‰ (*pers. comm.*), which are less anomalous than those of MS grains obtained in this study (Fig. 1) and thus also suggest Mo contamination.

Moreover, the fact that a higher fraction of AB1 grains (50%) in this study show *s*-process isotopic compositions with larger deviations from the solar composition relative to the AB2 (17%) grains from Liu et al. (2017b) provides further support for the division of the AB grains into two subgroups, in addition to the indications from their C, N, Si, and Al isotope ratios. Note that the Mo isotope data of 57 MS, Y, and Z grains located on the same sample mounts that were analyzed in the same analytical session as the AB1 and AB2 grains consistently indicate that up to 20% of the grains suffered from severe Mo contamination, resulting in normal measured Mo isotopic compositions (Liu et al. 2017b). Consequently, the normal Mo isotopic compositions of about three of the AB1 grains analyzed here might have been caused by contamination instead of progenitor stellar nucleosynthesis. In turn, it can be inferred that 11 (instead of 8) out of the 16 AB1 grains (69%) had *s*-process Mo isotopic compositions, in complete agreement with the result of Amari et al. (1995, 2001) that 65% of all of the 20 AB1 grains that they analyzed had *s*-process enhancements. Interestingly, grain M1-A7-G987 with the largest Mo isotopic anomalies also had large $^{32}$S and $^{50}$Ti enhancements, as shown by Liu et al. (2017a).

In order to compare with the AB1 grain data, we conducted new nucleosynthesis calculations based on the simplified SN model presented in Bojazi & Meyer (2014). This model computes SN shock passage by applying the observation of Woosley & Weaver (1995) that, during shock passage through the star, the explosion energy $E$ is approximately uniformly distributed in the shocked material and that the energy density is dominated by relativistic particles. The consequence of these assumptions is that the peak temperature a stellar layer reaches as the shock passes is proportional to $E^{1/4}$. The time evolution of the temperature in a layer after shock passage then depends on the shock speed outside that layer, which the model provides, while the density is computed through the local equation of state. In this way, the model permits one to compute thermodynamic trajectories for each stellar layer and then the resulting nucleosynthesis. For the present study, we computed the explosive nucleosynthesis for an initially 25 $M_{\odot}$ presupernova (pre-SN) model from Rauscher et al. (2002). The reaction network was the same as employed in Bojazi & Meyer (2014). The network can include all isotopes between the proton- and neutron-drip lines for species ranging from neutrons and protons up to the trans-uranic elements. For efficiency, the network includes only those species required at a given time step for accurate calculation of the abundances. The nuclear and reaction rate data were those from the JINA ReaclibV2.0 snapshot (Cyburt et al. 2010). We explored the nucleosynthesis resulting from explosion energies $E$ ranging from $0.75 \times 10^{51}$ erg to $5 \times 10^{51}$ erg. We did not include neutrino-nucleus interactions, since we focus on stellar layers well away from the collapsing stellar core. In these outer layers, the neutrino flux is too small to have a significant effect on the nucleosynthesis of interest.

## 4. DISCUSSION

### *4.1. Possible Stellar Sources of AB1 Grains*

As discussed by Liu et al. (2017a), the most likely stellar sources of AB1 grains are SNe that experienced explosive H burning in their outer layers, based on the fact that AB1 grains show both explosive H burning and *n*-capture isotopic signatures. In addition, the explosive H burning isotopic signatures of AB1 grains, such as elevated $^{13}C/^{12}C$, $^{15}N/^{14}N$, and inferred $^{26}Al/^{27}Al$ ratios, depend strongly on the peak temperature achieved and the amount of H present during the explosion (Liu et al. 2016, 2017a). The fact that AB1 grains define a tight correlation between their $^{14}N/^{15}N$ and inferred $^{26}Al/^{27}Al$ ratios strongly indicates that they share a common stellar origin, namely SNe with explosive H burning. In contrast, another suggested stellar source for AB grains, born-again AGB stars like Sakurai's object (Fujiya et al. 2013), are less likely, based on the fact that the state-of-the-art models for Sakurai's object predict lower-than-solar $^{15}N/^{14}N$ ratios for the $^{13}C$-rich region of a born-again AGB star (Herwig et al. 2011), while, in contrast, AB1 grains are characterized by $^{15}N/^{14}N$ ratios in excess of the solar value. However, it is noteworthy that the H entrainment rate during the post-AGB phase depends on the initial progenitor stellar mass, which can lead to quite different N isotope ratios in the $^{13}C$-rich region of a post-AGB star, and more modeling efforts are therefore needed for comparison with presolar grain data.

The heavy-element isotope data provide further evidence against a born-again AGB stellar origin of AB1 grains. The model calculations by Herwig et al. (2011) for Sakurai's object predict that the intermediate *n*-capture process (*i*-process, Cowan & Rose 1977) takes place in born-again AGB stars at enhanced neutron densities (~$10^{15}$ cm$^{-1}$) relative to those for the *s*-process (~$10^{7}$ cm$^{-1}$) in AGB stars, resulting in an anomalous pattern of elemental abundances (relative to *s*-process elemental patterns) observed on the surface of the Sakurai's object (Asplund et al. 1999). Specifically, the Herwig et al. (2011) models predict extremely anomalous Sr, Mo, and Ba isotopic compositions for born-again AGB stars (model predictions for Ba isotopes are shown in Fig. 10 of Liu et al. 2014, and the corresponding predictions for Mo isotopes are given in Fig. 1b), e.g., large positive $\delta^{97,98,100}$Mo values (Fig. 1b), which are inconsistent with the *s*-process isotopic signatures seen in the AB1 grains from this study. Thus, an origin in born-again AGB stars seems unlikely, but the effect of the initial progenitor stellar mass on Mo isotope production in post-AGB stars needs to be investigated for comparison with the grain data. In the following sections, we will discuss the Sr, Mo, and Ba isotopic signatures of AB1 grains in the context of SN nucleosynthesis.

*4.2. Lack of Neutron-Burst Signatures in AB1 Grains*

In Fig. 2, we show the model predictions for the He/C and He/N zones of a 25 $M_{\odot}$, $Z_{\odot}$ (Anders & Grevesse 1989) SN with two vastly different explosion energies. We have adopted the simulations of Rauscher et al. (2002) for the pre-SN stellar compositions. For reference, Rauscher et al. (2002) adopted an explosion energy of $10^{51}$ erg for the same pre-SN stellar model in their calculations. It is obvious from Fig. 2 that the bottom of the He/C zone (at 7 $M_{\odot}$) is characterized by large excesses in the most neutron-rich isotopes, including $^{50}$Ti, $^{88}$Sr, $^{100}$Mo, and $^{138}$Ba as a result of the so-called neutron-burst process powered by the $^{22}$Ne($\alpha$,$n$)$^{25}$Mg reaction during SN explosions (Meyer et al. 2000). In contrast, the outer region of the He/C zone (shaded area in Fig. 2) remains unmodified by the neutron-burst process, and preserves the pre-SN *n*-capture signatures, because the peak temperature reached during the explosion is too low to activate the $^{22}$Ne($\alpha$,$n$)$^{25}$Mg reaction in this region. With increasing explosion energy, the neutron-burst process occurs more efficiently, resulting in more anomalous isotopic signatures at the bottom of the He/C zone and affects the nucleosynthetic signatures of a larger fraction of the He/C zone. Explosive *n*-capture isotopic signatures were previously found in the main X subgroup, X1 grains (Pellin et al. 2000; Stephan et al. 2018) and also in one of the five AB1 grains studied by Savina et al. (2003). However, none of the AB1 grains in this study showed such neutron-burst isotopic signatures. The isotopic compositions of the isotopically anomalous AB1 grains are distinctly different from the predictions for the explosive *n*-capture process (Fig. 2), and are more similar to the pre-SN signatures predicted by the model calculations except for $\delta^{49}$Ti (Fig. 1c).

Pignatari et al. (2018, hereafter P18) recently conducted a detailed analysis of *n*-capture nucleosynthesis in SN outer zones in the context of Zr and Mo isotopic signatures of presolar X1 grains. The model calculations shown in Fig. 2 agree generally with the P18 model calculations for *s*-process elements, except for the fact that we did not find the formation of a Si/C zone even by increasing the explosion energy to $5\times10^{51}$ erg. In contrast, Pignatari et al. (2015, hereafter P15) were able to produce such a Si/C zone at the bottom of the He/C zone by adopting explosive energies of $(4-7)\times10^{51}$ erg for a 25 $M_{\odot}$ SN. This discrepancy is related to differences in the details of the pre-SN evolutionary simulations adopted in the two sets of models. Nevertheless, our calculations confirm the general features of the P18 models (e.g., Fig. 3 of P18) that the inner He/C zone is significantly modified by the neutron-burst process whilst the outer region (shaded area in Fig. 2, zone 4 of P18) remains unmodified. In addition, similar to the P18 models, our models also predict large $^{100}$Mo excesses and higher enrichments in $^{94}$Mo relative to $^{92}$Mo for the neutron-burst process, which are not observed in X1 SiC grains that sampled neutron-burst materials from their parent SNe (Pellin et al. 2000) and thus imply problems related to the productions of $^{94}$Mo and $^{100}$Mo in the current SN model calculations, as pointed out by P18.

*4.3. Pre-SN n-Capture Environment*

P15 investigated the effect of H ingestion (up to 1.2% by mass) into the He/C zone in the pre-SN phase, on explosive nucleosynthetic isotopic signatures during SN explosions (see P15 for model details). The P15 models (25T and 25d models) showed that large amounts of $^{13}$C, $^{15}$N, and $^{26}$Al are produced as a result of explosive H burning at the bottom of the He/C zone. Based on the P15 models, Liu et al. (2017a) were able to reproduce the $^{14}$N/$^{15}$N and $^{26}$Al/$^{27}$Al correlation seen in AB1 grains.

The pre-SN *n*-capture isotopic compositions of AB1 grains confirm the P15 model prediction that the presence of H at the bottom of the He/C zone during SN explosions suppresses the neutron-burst process there. We also further tested this effect by adding 1% H in the He/C zone in our pre-SN models, and verified the result of P15 that this amount of H is sufficient to completely deactivate the neutron-burst process in the He/C zone during SN explosions in our models by adopting energies from $0.75\times10^{51}$ erg to $5\times10^{51}$ erg. In turn, compared to the He/C zone with varying neutron-capture isotopic signatures in Fig. 2, the H-ingested He/C zone in all the models after SN explosions exhibits the pre-SN isotopic signature (Fig. 1c) across the zone. This suppression of the explosive *n*-capture process is due to the $^{22}$Ne($\alpha$,*n*)$^{25}$Mg reaction becoming deactivated as a result of the competing $^{22}$Ne(*p*,$\gamma$)$^{23}$Na reaction powered by protons (P15). Thus, material in the He/C zone that experiences explosive H burning is naturally expected to show pre-SN *n*-capture isotopic signatures, as seen in AB1 grains.

Heterogeneous H mixing was previously proposed to explain the AB1 grain data, based on the observation that $^{32}$S and $^{50}$Ti excesses and explosive H burning signatures were concomitantly incorporated into AB1 grains (Liu et al. 2016, 2017a). This is mainly because current SN models predict that $^{32}$S and $^{50}$Ti can only be abundantly produced by the neutron-burst process at the bottom of the He/C zone during SN explosions and that the neutron-burst process becomes completely suppressed by the presence of H during SN explosions. Since the overall Sr, Mo, and Ba isotopic patterns of the AB1 grains are consistent with pre-SN rather than explosive, *n*-capture isotopic signatures, the mismatch of the current SN model predictions with the grain data points to a problem in the current model predictions for the pre-SN *n*-capture process. In turn, the coexisting pre-SN *n*-capture and explosive H burning isotopic signatures of AB1 grains can be well explained by the P15 models *as a natural consequence of explosive H burning without invoking heterogeneous H mixing*.

Specifically, the AB1 grain data suggest that the pre-SN *n*-capture process occurred with enhanced neutron densities with respect to the current model predictions. Specifically, the lower $^{49}$Ti/$^{50}$Ti ratios of the AB1 grains compared to the pre-SN *n*-capture model predictions (Fig. 1c; also see Liu et al. 2017a) cannot be due to errors in *n*-capture cross sections, since these for $^{49}$Ti and $^{50}$Ti are known to ~10% accuracy (Karlsruhe Astrophysical Database of Nucleosynthesis in Stars version v1.0, Dillmann et al. 2014)[11]. Rather, the data suggest that the pre-SN *n*-capture process occurred with enhanced neutron densities with respect to the current model predictions, since the predicted $^{50}$Ti/$^{49}$Ti ratio increases with increasing neutron density (e.g., Zinner et al. 2007). Moreover, the most anomalous Mo isotopic pattern observed in an AB1 grain, M1-A7-

[11] Available online at http://exp-astro.physik.uni-frankfurt.de/kadonis/index.php.

G987, cannot be explained by the current model predictions nor can this grain's large $^{32}$S and $^{50}$Ti excesses. It is also worth mentioning that although most of the AB1 grain data can be matched by the current pre-SN model predictions in Fig. 1c, mixing of H-envelope material that has normal Ti, Sr, Mo, and Ba isotopic compositions with explosive H burning product from the He/C zone is required to explain the Al-N isotope trend observed in AB1 grains (Liu et al. 2017a). This is likely to further dilute the *n*-process isotopic signatures in the precursor SN mixtures of AB1 grains, and a stronger *n*-capture process is thus generally required to explain the AB1 grain data.

### *4.4. Explosive H Burning in the He/N Zone*

Our calculations show that, in addition to the bottom of the He/C zone, explosive H burning can also occur at the top of the He/N zone at relatively low temperatures, corresponding to the $^{15}$N-rich spike previously reported by Nittler (2009) and Lin et al. (2010). This burning occurs as a result of pre-SN mixing of a large amount of H (up to 40% by mass, Fig. 3a) from the adjacent H envelope in the pre-SN phase (Fig. 3) due to semiconvection and/or overshoot (Canuto 2000). In the $5\times10^{51}$ erg case, the $^{12}$C/$^{13}$C and $^{14}$N/$^{15}$N ratios are predicted to reach minima of 1.9 and 0.01, respectively, and thus can well account for the C and N isotope ratios of all AB1 grains and most of the putative nova grains in the literature (e.g., Liu et al. 2016, 2017a). Figure 4 further illustrates the trends of different ratios varying with increasing explosion energy, and, of these, the $^{14}$N/$^{15}$N ratio shows the strongest dependence on the explosion energy. Both the P15 H-ingestion models and the model predictions shown here strongly indicate that explosive H burning can occur at multiple locations within the He/N and He/C zones depending on the H mixing mechanism and the peak temperature achieved during a SN explosion. Also, the predicted N and Al isotope ratios in Fig. 4 are similar to the explosive H burning endmember inferred by the N-Al isotope trend of AB1 grains (e.g., Fig. 1 of Liu et al. 2017a), which points to the general similarity of explosive H burning products (He/N versus He/C) within SNe. Since the pre-SN isotopic signatures of heavy elements are predicted to be normal in the He/N zone, the normal Mo isotopic composition of the ~30% "uncontaminated" AB1 grains can also be explained. Note that if explosive H burning only occurred in the He/N zone of the progenitor star of a SN SiC grain, the He/N zone material alone probably cannot provide a C-rich environment for SiC grains to condense, because, even in the $5\times10^{51}$ erg case, the C/O ratio of the He/N zone barely reaches unity (Fig. 4). In addition, mixing with the extremely O-rich material from the H envelope as required to explain the N-Al isotope trend of AB1 grains is likely to further reduce the C/O ratio. In this case, material from the He/C zone needs to be mixed with the He/N zone material to raise the C/O ratio. Thus, if AB1 grains with normal heavy-element isotopic compositions inherited their H-ingestion signatures from the He/N zone, it requires their parent SNe to have low explosion energies (e.g., less than $5\times10^{51}$ erg for a 25 $M_\odot$ progenitor star) so that normal heavy-element isotopic composition can be preserved in the outer part of the He/C zone (e.g., Fig. 2).

### *4.5. Further Evidence of Explosive H Burning in SNe*

The occurrence of explosive H burning in SNe is also supported by the isotopic compositions of X grains. Previous studies have shown that explosive H burning material from the $^{15}$N-rich spike in the He/N zone and/or from the H-ingested region in the He/C zone is required to explain the C and N isotopic ratios of X grains (Lin et al. 2010; Hoppe et al. 2018). Thus, SN X grains suggest that explosive H burning occurred in the outer C-rich He/C and/or He/N zones of their parent SNe, as also argued for the progenitor SNe of C2 and AB1 grains (Liu et al. 2016, 2017a), implying that it is a common SN phenomenon.

Evidence that X1 and X2 grains have different *n*-capture isotopic signatures has been recently reported. While X1 grains exhibit clear neutron-burst signatures (Pellin et al. 2000; Stephan et al. 2018), pre-SN signatures were found in the only two X2 grains that have been studied for their heavy-element isotopic compositions so far (Stephan et al. 2018). As mentioned above, the C and N isotope ratios of X grains strongly suggest that explosive H burning generally occurs in the SN C-rich region. Thus, the pre-SN *n*-capture isotopic compositions of both X2 and AB1 grains are consistent with the P15 model prediction that explosive H burning deactivates the neutron-burst process in the He/C zone. The presence of neutron-burst signatures in X1 grains, however, contradicts this picture. While heterogeneous H ingestion into the He/C zone is one possible explanation, an alternative is that in the parent SNe of X1 grains, explosive H burning mainly occurred in the He/N zone, instead of in the He/C zone where the neutron-burst process takes place.

## 5. CONCLUSION

We obtained Mo isotopic ratios in 16 AB1 grains, with correlated Sr and Ba isotope data in subsets of these grains. Direct comparison of the AB1 grains with AB2 and MS grains from Liu et al. (2017b) shows that AB1 grains have *s*-process Mo isotopic signatures that are stronger than those of AB2 grains, providing another piece of evidence to support the division of the two subgroups. Although born-again AGB stars have been suggested as a possible source for AB1 grains, the state-of-the-art born-again AGB models for the Sakurai's object cannot explain the fact that the *s*-process Mo isotopic signatures of AB1 grains are weaker than those of MS grains.

The most likely stellar sources of AB1 grains are SNe that experienced some explosive H burning in their He/C and/or He/N zones. In addition, comparison of a new set of SN model predictions with the AB1 grain data for Sr, Mo, and Ba isotopes indicates that the *s*-process isotopic compositions of AB1 grains reflect pre-SN, rather than explosive, *n*-capture processes. Thus, the Sr, Mo, and Ba isotope data of AB1 grains clearly show that *no heterogeneous H mixing is needed to explain the grain data, as some have invoked,* and the mismatch of the SN model predictions for $^{49}$Ti/$^{50}$Ti with the AB1 grain data suggests that pre-SN *n*-capture process occurred with enhanced neutron densities with respect to the current model predictions.

As predicted by stellar model calculations (P15), H (up to 1.2% by mass) can be ingested into the He/C zone in the pre-SN phase, resulting in a significant production of $^{13}$C and lowered $^{12}$C/$^{13}$C ratio in the He/C zone during SN explosions. On the other hand, our new calculations show that explosive H burning can also occur in the He/N zone at relatively low temperatures, as a result of inward mixing of a large amount of H from the adjacent H envelope. In support of this, previous studies have shown that explosive H burning in the He/N zone and/or the He/C zone is required to explain X grain isotopic data (Lin et al. 2010; Hoppe et al. 2018). The $^{13}$C-enhanced environment for the outer C-rich region (including both the He/C and He/N zones in a SN) suggested by both X and AB1 grains is thus naturally expected according to SN model calculations. More modeling is needed in the future to systematically investigate the effects of different mixing mechanisms, e.g., magnetic buoyancy, in the pre-SN phase on the amount of H mixed into both the He/C and He/N zones and the corresponding distributions of the mixed H.

Acknowledgements: This work was supported by NASA (grants NNX10AI63G and NNX17AE28G to LRN, NNX15AF78G and 80NSSC17K0250 to AMD, and NNX14AI25G to BSM). We thank Jianhua Wang for technical support on the NanoSIMS, and Falk Herwig and Marco Pignatari for providing post-AGB model predictions for Mo isotopes, and Michael Savina for sharing the N isotope data of seven AB grains and Mo isotope data of MS grains from Savina et al. (2003) that were not reported in the original abstract. Part of this work was performed under the auspices of the U.S. Department of Energy by Lawrence Livermore National Laboratory under Contract DE-AC52-07NA27344 and by the Laboratory Directed Research and Develop- ment Program at LLNL under project 17-ERD-001. LLNL-JRNL-735278

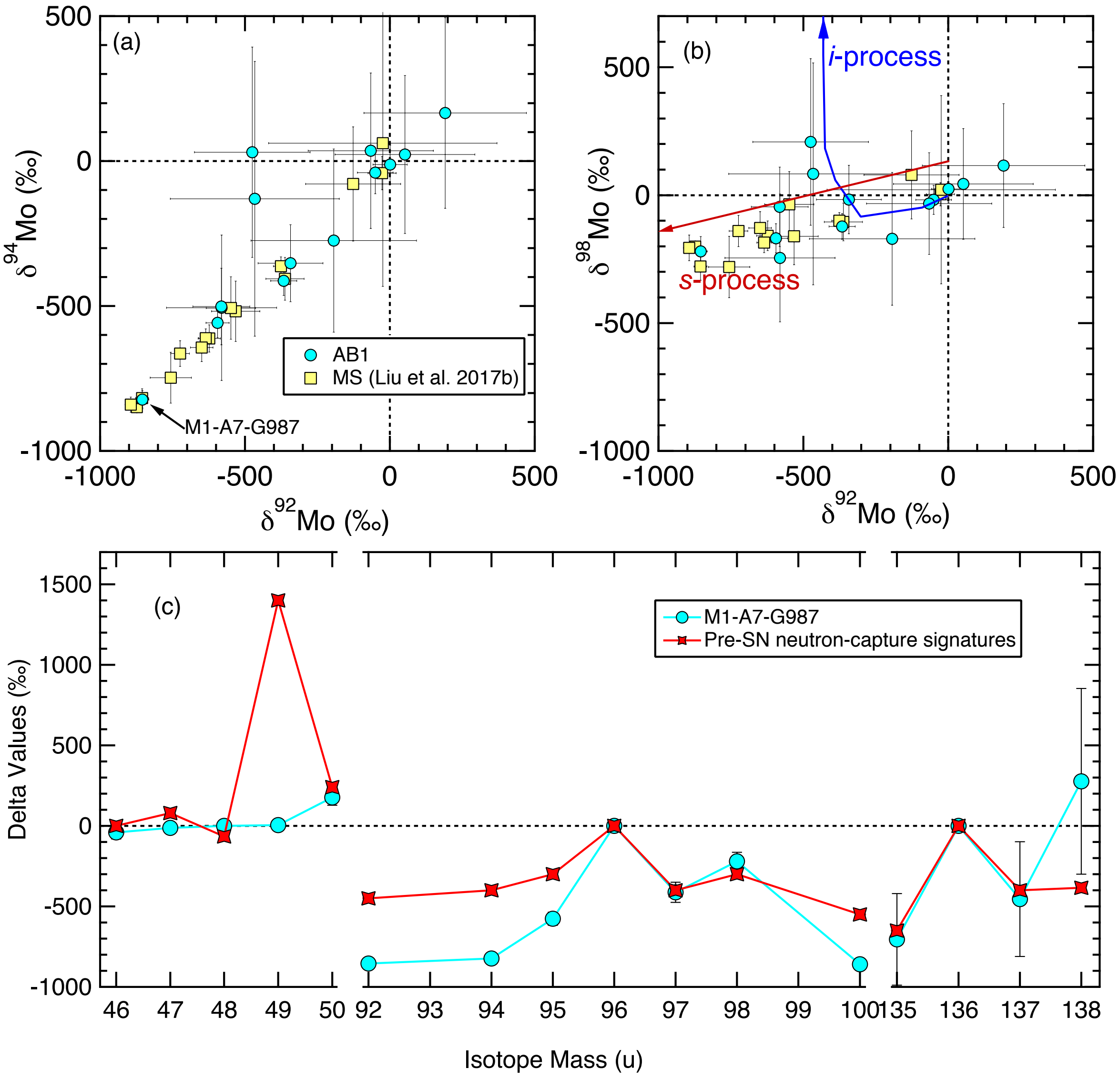

**Fig. 1.** Molybdenum three-isotope plots of $\delta^{94}$Mo (a) and $\delta^{98}$Mo (b) versus $\delta^{92}$Mo for SiC AB1 grains from this study and MS grains from Liu et al. (2017b). All the grains were on the same sample mounts and analyzed in the same analytical session with CHILI. Two AB1 grains with 2σ errors larger than 400‰ in Table 1 are not shown in the plot for clarity. The *s*-process endmember is indicated by the red arrow in (b) for comparison. For the *i*-process, the weight-averaged Mo isotopic composition of a helium zone ingested by hydrogen after a VLTP during the post-AGB phase (Herwig et al. 2011, a split time of 1000 min) is shown as a blue line for comparison; the final composition is $\delta^{92}$Mo=−1000‰, $\delta^{98}$Mo=3050‰. (c) The Ti, Mo, and Ba isotopic patterns of grain M1-A7-G987 (stable isotopes are indicated by symbols), the AB1 grain with the most anomalous Mo isotopic composition, in comparison to the pre-SN isotopic signatures in the shaded area of Fig. 2. All the grain data are plotted with 2σ errors.

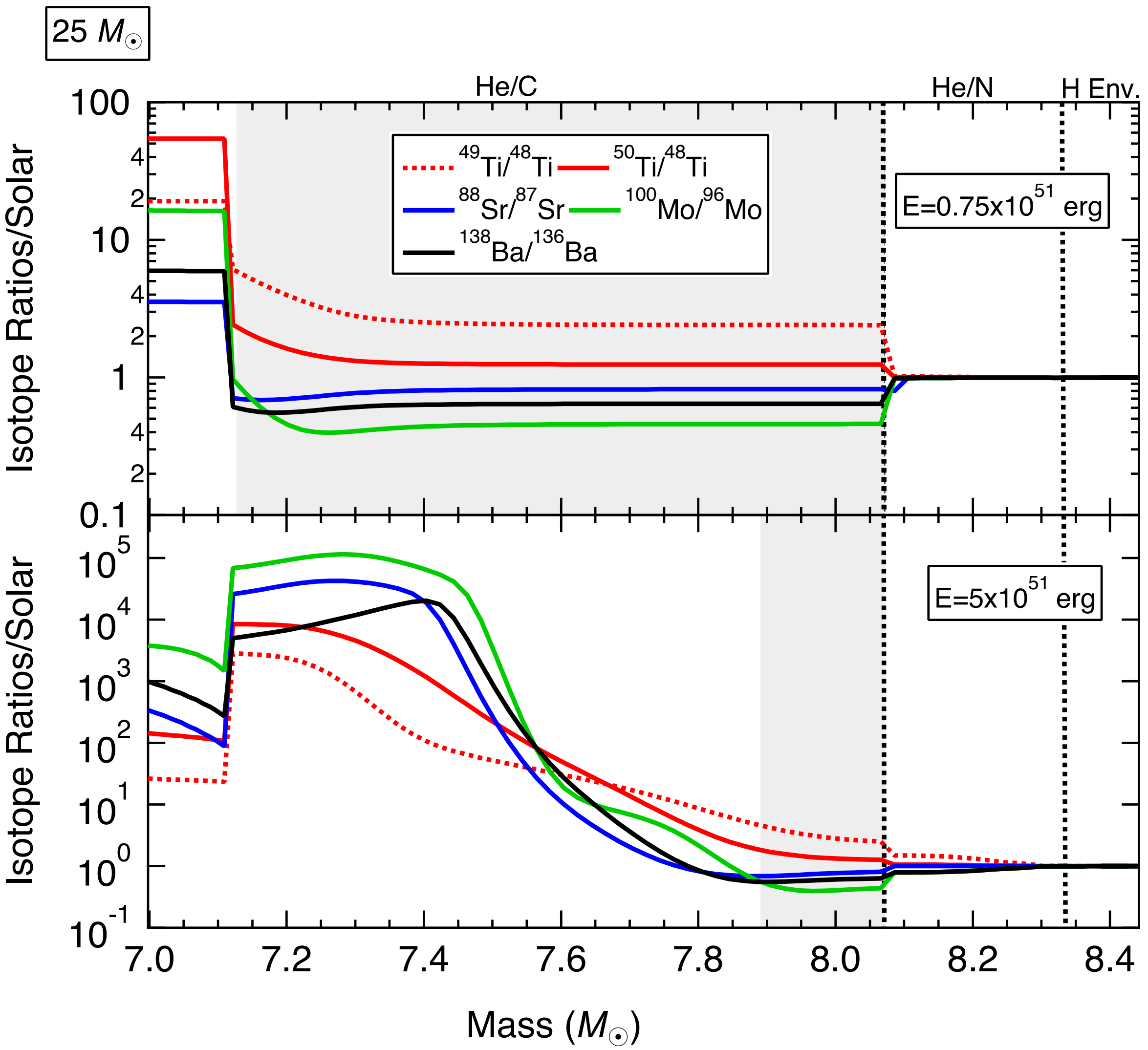

**Fig. 2.** Model predictions for the He/C and He/N zones and part of the H envelope of a 25 $M_\odot$, $Z_\odot$ SN with two vastly different explosive energies based on the pre-SN stellar evolutionary simulation of Rauscher et al. (2002). The isotopic ratios shown in the two plots are atomic ratios normalized to their corresponding solar values, so a predicted isotope ratio of unity corresponds to its solar value. The shaded areas highlight the parts of the He/C zones that are unaffected by the explosive *n*-capture process and preserve the pre-SN *n*-capture isotopic signatures.

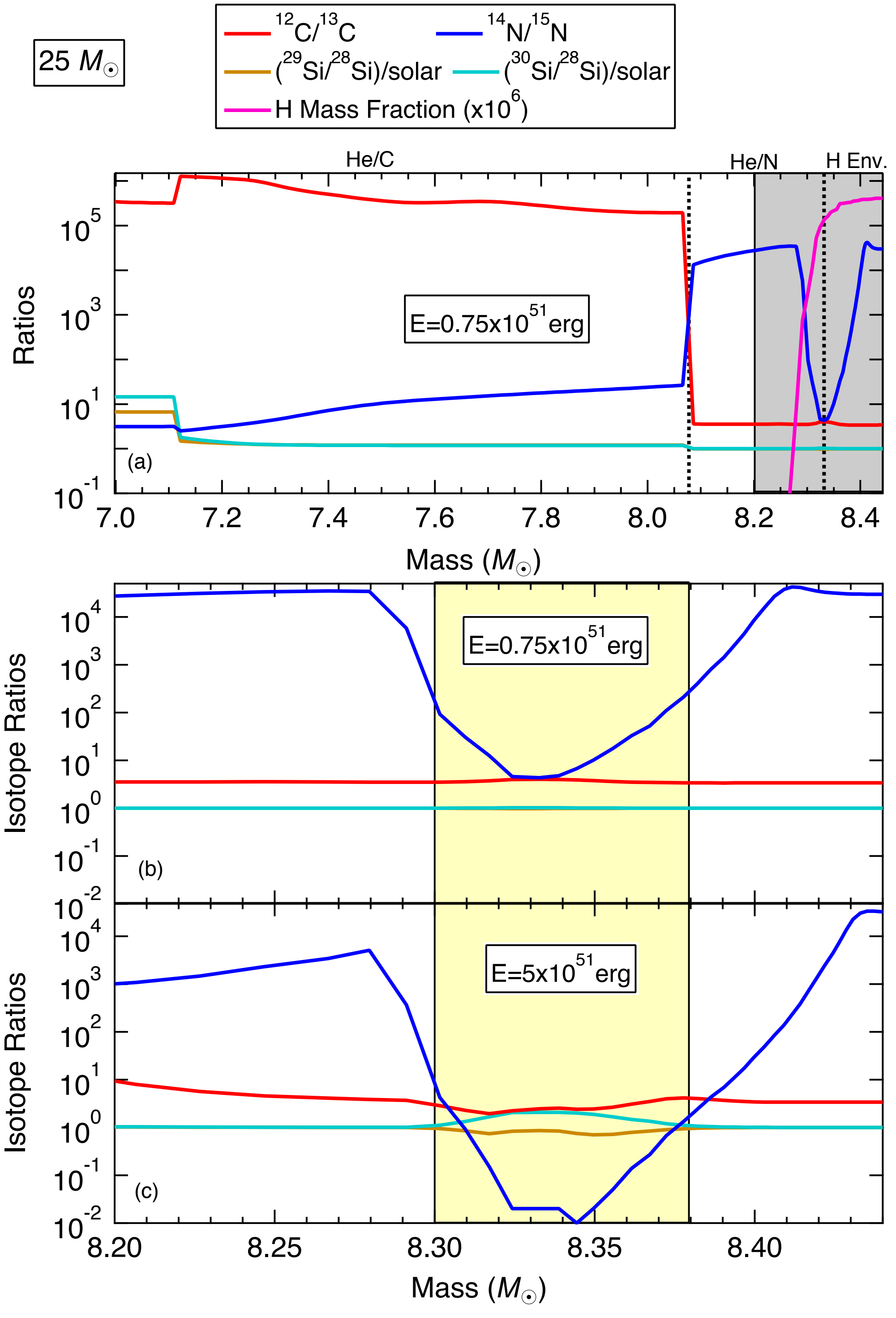

**Fig. 3.** (a) Model predictions in the same region as in Figs. 2. (b) & (c) Zoom-in plots of the shaded area in (a). The yellow area highlights the region where explosive H burning occurs during a SN explosion.

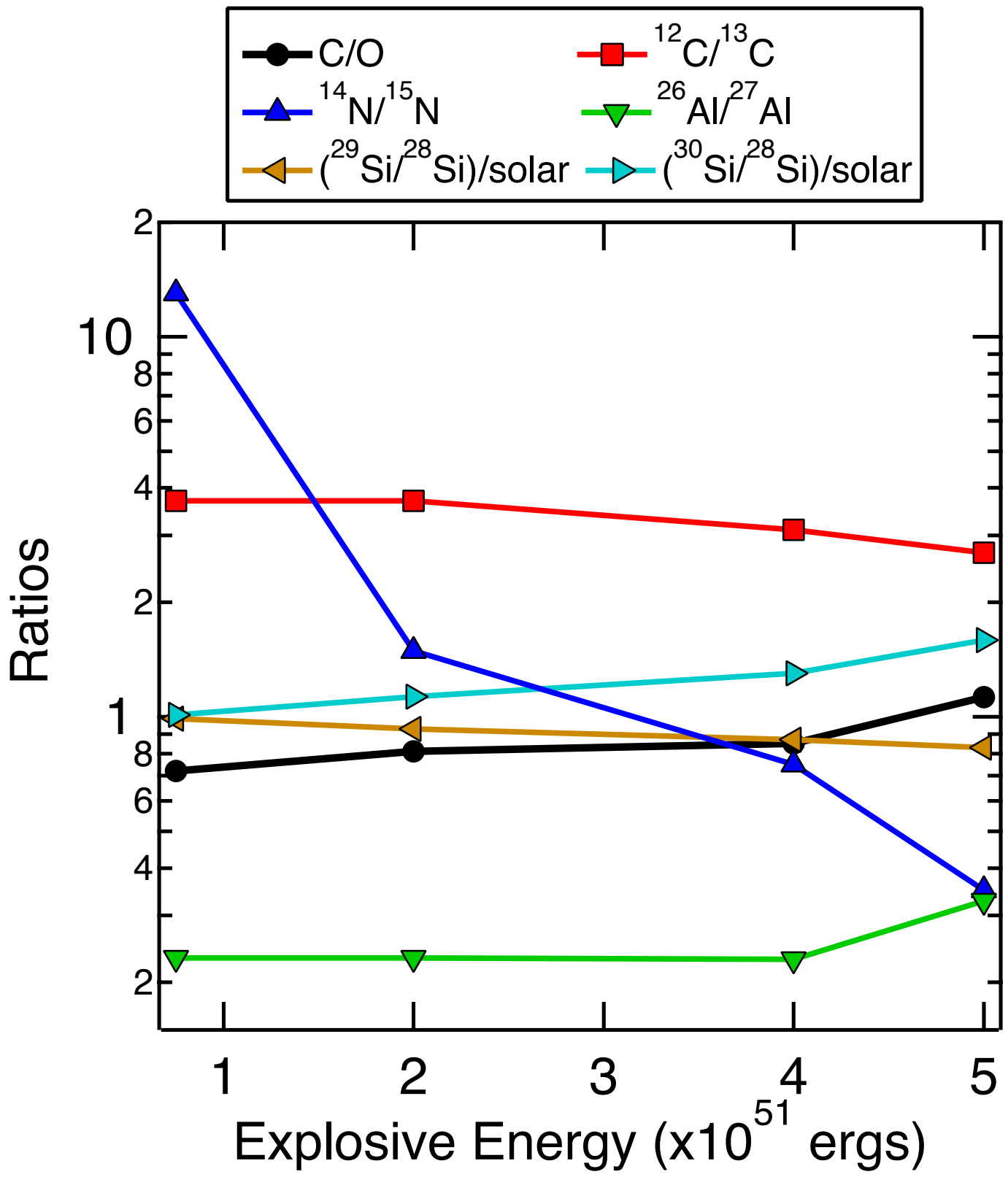

**Fig. 4.** Averaged ratios in the yellow area in Figs. 3b,c versus the explosion energy.

**Table 1.** Isotopic Data of AB1 Grains.

| Grain | Group | Size ($\mu m^2$) | $^{12}C/^{13}C$ | $^{14}N/^{15}N$ | $\delta^{29}Si$ (‰) | $\delta^{30}Si$ (‰) | $\delta^{92}Mo$ (‰) | $\delta^{94}Mo$ (‰) | $\delta^{95}Mo$ (‰) | $\delta^{97}Mo$ (‰) | $\delta^{98}Mo$ (‰) | $\delta^{100}Mo$ (‰) | $\delta^{84}Sr$ (‰) | $\delta^{88}Sr$ (‰) | $\delta^{135}Ba$ (‰) | $\delta^{137}Ba$ (‰) | $\delta^{138}Ba$ (‰) |
|---|---|---|---|---|---|---|---|---|---|---|---|---|---|---|---|---|---|
| M1-A4-G386 | AB1 | 1.3×1.0 | 6.3±0.02 | 100±4 | 8±46 | −29±46 | 191±280 | 165±329 | 226±289 | 184±317 | 116±241 | 80±306 | | | | | |
| M1-A4-G428 | AB1 | 1.5×1.5 | 4.1±0.04 | 195±18 | 70±52 | 24±52 | 52±241 | 23±273 | 195±261 | 55±278 | 44±216 | 129±290 | | | | | |
| M1-A4-G557 | AB1 | 1.0×0.8 | 3.5±0.02 | 144±8 | 33±32 | 39±22 | −343±112 | −352±133 | −185±130 | −177±150 | −17±134 | −306±137 | | | −828±305 | −483±501 | −391±425 |
| M1-A4-G574 | AB1 | 1.5×0.8 | 4.7±0.02 | 261±22 | −40±44 | −24±46 | −366±44 | −413±50 | −247±48 | −201±62 | −122±49 | −401±50 | | 123±341 | | | |
| M1-A5-G890 | AB1 | 0.6×0.6 | 6.1±0.06 | 386±56 | −26±42 | 3±44 | −581±189 | −506±251 | −444±223 | −454±270 | −245±250 | −528±251 | | | | | |
| M1-A6-G594 | AB1 | 0.8×0.7 | 4.4±0.04 | 252±30 | 89±64 | 68±70 | −195±286 | −274±316 | −144±293 | −239±327 | −171±260 | −16±386 | | | | | |
| M1-A7-G866 | AB1 | 0.6×0.6 | 6.1±0.10 | 387±112 | −13±64 | 39±78 | −595±39 | −558±50 | −391±50 | −306±65 | −169±57 | −597±48 | | | −797±341 | −493±505 | 38±697 |
| M1-A7-G987 | AB1 | 1.0×1.3 | 5.0±0.04 | 384±34 | −17±40 | 23±40 | −854±23 | −823±31 | −577±41 | −412±62 | −220±57 | −859±28 | | | −705±285 | −454±356 | 276±576 |
| M2-A1-G513 | AB1 | 1.0×1.0 | 4.5±0.26 | 252±34 | 11±20 | 30±20 | −475±199 | 30±362 | −44±295 | −288±286 | 208±326 | −401±259 | | | | | |
| M2-A2-G609 | AB1 | 1.1×1.0 | 2.8±0.12 | 129±14 | −23±24 | −17±30 | −150±536 | 155±776 | −245±480 | 471±898 | −67±520 | −92±640 | | | | | |
| M2-A2-G722 | AB1 | 0.7×0.7 | 4.1±0.18 | 100±10 | 20±28 | 40±36 | −348±435 | −295±538 | −58±561 | −119±611 | 202±619 | −456±452 | | | | | |
| M2-A2-G730 | AB1 | 1.0×0.8 | 5.8±0.26 | 192±22 | 31±22 | 13±26 | −466±291 | −130±473 | −167±410 | 96±543 | 83±434 | −674±294 | | | | | |
| M2-A4-G790 | AB1 | 1.0×1.0 | 7.6±0.11 | 89±14 | −1±26 | 28±28 | −582±98 | −502±132 | −337±130 | −257±168 | −46±155 | −607±117 | | 81±303 | −993±474 | −399±517 | −302±464 |
| M3-G587 | AB1 | 1.0×0.8 | 4.8±0.04 | 300±34 | 171±14 | 141±18 | −51±62 | −40±72 | −19±62 | 2±73 | −18±57 | −3±73 | | −99±185 | | | |
| M3-G1485 | AB1 | 1.0×0.6 | 3.0±0.01 | 102±6 | −39±22 | −44±28 | −66±215 | 35±268 | 49±230 | 64±272 | −33±199 | 27±266 | | | | | |
| M3-G1718 | AB1 | 0.8×0.6 | 4.2±0.02 | 166±12 | 47±24 | 48±26 | 1±58 | −12±66 | −26±55 | 25±67 | 24±53 | −31±65 | −431±376 | −372±87 | | | |

Note: δ-notation is defined as $\delta^{i}A=[(^{i}A/^{j}A)_{grain}/(^{i}A/^{j}A)_{std}-1]\times1000$, where A denotes an element, i an isotope of this element, and j the normalization isotope, and $(^{i}A/^{j}A)_{grain}$ and $(^{i}A/^{j}A)_{std}$ represent the corresponding isotope ratios measured in a sample and the standard, respectively. The normalization isotopes are $^{28}Si$, $^{87}Sr$, $^{96}Mo$, and $^{136}Ba$. All values are reported with 2σ errors.